## Implementation of Automata Theory to Improve the Learning Disability

S. A. ALI[++], S. SOOMRO, A. G. MEMON* A. BAQI**

Department of Computer Science, Faculty of Information Technology, Sindh Madressatul Islam University, Karachi, Pakistan

**Abstract:** There are various types of disability egress in world like blindness, deafness, and Physical disabilities. It is quite difficult to deal with people with disability. Learning disability (LD) is types of disability totally different from general disability. To deal children with learning disability is difficult for both parents and teacher. As parent deal with only single child so it bit easy. But teacher deals with different students at a time so it's more difficult to deal with group of students with learning disability.
If there is more students with learning disability so it is necessary that first all identify the type of learning disability in group of students. Some students have learning disability of mathematics; some have learning disability of other subjects. By using theory of Automata it easy to analysis the level of disability among all students then deal with them accordingly. For these purpose deterministic automata is the best practice. Teacher deals with deterministic students in class and check there response. In this research deterministic automata is use to facilitated the teacher which help teacher in identification of students with learning disability.

## I. INTRODUCTION

Teaching is an art, it is important to know how to perform this art. Teacher must be aware how to deals with their students. Here these points should be considered that if the teacher deals with a group of students who have learning disability than must consider the methods that how to attend this group of students. Learning disability is not a big problem and can easily be handling by making few new teaching strategies, like more positive toward students and create emotional class room environment. The deterministic finite automata are best practice to form emotional class room environment for students with learning disabilities.

## 2. MATERIAL AND METHODS

**Deterministic Finite Automata**

A deterministic finite automata (DFA) is a quintuple M= (Q, $\sum$, $\delta$, $q_0$, F), where Q is a finite set of states, $\sum$ a finite set called alphabet, $q_0 \in Q$, a distinguished state known as the start state, F a subset of Q called the final or accepting states, and $\delta: Q \times \sum \to Q$, known as the transition function.

The extended transition function can be defined. Let M= (Q, $\sum$, $\delta$, $q_0$, F) be a DFA, we define the function $\delta^*: Q X \sum^* \to Q$ as follows:
1. For any q$\in Q$, $\delta^*(q, \lambda) = q$.
2. For any q $\in Q$, $y \in \sum^* and\ a \in \sum$ then $\delta^*(q, ya) = \delta(\delta^*(q, ya))$.

A string is accepted by M if $\delta^*(q_0, x) \in F$. Thus the language recognized by the DFA M is the set we have referred to as a deterministic finite automata as an abstract machine. The operation of a DFA is described in terms of components that are present in many familiar computing machines.

A computation of an automaton consists of the execution of a sequence of instructions where the execution of an instruction alters the state of the machine to some new state. The objective of a computation of an automaton is to determine the acceptability of the input string. An input string is accepted if the computation terminates in an accepting state; otherwise it is rejected. At any point during the computation, the result depends only on the current state and the unprocessed input. This combination is called a machine configuration and is represented by the ordered pair [$q_i$, w], where qi is the current state and $w \in \sum^*$ (Hopcroft *et al.*, 2002), and (Hu *et al.*, 1996).

**Learning Disability (LD)**

The term learning disability is used to express specific category of learning deficiency .The persons with learning disability face lot of problems during learning process in certain skills. The most common skills are reading, writing, listening, speaking, interpretation etc.

++ Corresponding author: Email: aasyed@smiu.edu.pk, ssoomro@smiu.edu.pk
*Institute of Mathematics and Computer Science, University of Sindh
**Department of Computer Engineering, Salman bin Abdulaziz University Kingdom of Saudi Arabia

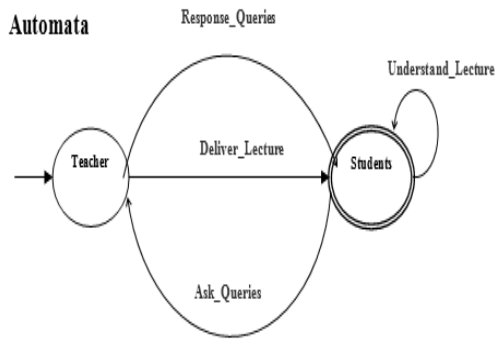

**Fig. 1: Automata for Simple Class room Environment**

Learning disabilities are originated by distinction in how a person's brain works. it different from persons to person may be one persons has Learning Disability in speaking and another person's has learning disability of reading (Smith, 2010), ( Harwell *et al.,* 2008 ).

Children with learning disabilities are not like blind or deaf persons. In reality, they typically have average or above average aptitude. Their brains just process information in a different way.

There is no treatment for learning disabilities. They are life-long. However, learning disability can cover in children by using teaching tips and special methodology so they can achieve their target and can and do learn successfully.

The quality of good teacher to identify the children with learning disability in classroom. Teacher should use common language and try to support for improving the quality of children and teacher relationship, which help in grooming the children with learning disability (Killen,1998), (Neisser, 1982).

## 3. DISCUSSION AND RESULT
**Enforcement of Deterministic Finite Automata on students with Learning Disability**

Enforcement of Deterministic Finite Automata is a best model to show teacher and students interaction in classroom environment, to overcome the problem of children specially children with learning disability.

This model support both emotional and instructional to children with learning disability.

**Language**
L= {Deliver_Lecture, Response_Queries, Understand_Lecture, Ask_Queries}

**Regular Expression**
r.e= (Deliver_Lecture))$^*$. Understand_Lecture +(Deliver_Lecture.(Ask_Queries.Resonse_Queries)$^*$)$^*$. Understand_Lecture

Teacher will deliver lecture notes and observed the response of the students. After getting response experienced teacher could easily identify the weak student of their class. But the weak students do not mean that these are students with learning disabilities. Learning disability will be observed by teacher after conducting few lectures. Teacher will apply the same model using emotional support, and asking questions from students if there are no queries from their side to check the response.

By the learning model teacher can firstly identify the children with learning disability and secondly check the class of learning disability either reading or writing etc. the model is not only successful for teacher, it is also best for their parent or education facilitator (Neisser, 1982)

The emotional class room environment is very important to create students interest especially for the students with learning disability, because they want special attention. On other hand self-control is an ability of good teacher but the boundary of self-control should be relaxed for students with learning disability .If teacher provides isolated environment then the students with learning disabilities feel more stress and the graph of learning disability will increase. They feel that they have no self-respect. Therefore the behaviors of teacher always are positive and more positive in case of learning disabled students.

Use of positive language also motivates the students. When emotional educational environment is established student with learning disability will show their positive response and resultant teacher earn respect with them and the ratio of asking question automatically reduce in such a environment.

The automata for emotional class room environment will track the teacher language ($L_1$) and the student's response ($L_2$).

**Teacher_Language**
$L_1$ ={Emotional Environment, Positive behavior, Positive Language, Extra motivation}

**Students_Response**
$L_2$={Understand_Lecture, Minimum_Queries, give_respect}

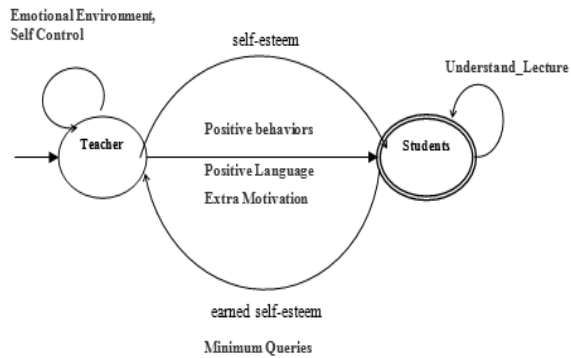

**Fig. 2: Automata for Emotional Class room Environment**

**Regular Expression**
r.e= (Deliver_Lecture.(Emotional Environment Positive behavior .Positive Language. Extra Motivation)))$^*$.Understand_Lecture .(Minimum_Queries.give_respect)$^*$)$^*$. Understand_Lecture.give_respect)

## 4. CONCLUSION

Theory of Automat is a best practice to identifying children with learning disabilities. Using Automata concept one can easily classify the type of learning disability too. Computational science is a matter of fact and indeed, it is a reality that evens today it is difficult for teacher to identify or classify the student with learning disability. More or less the entire learning model used in machine or automata theory. In common, Children learning system is directly associated with the teacher. The teacher style of different teachers of same subject may be different. Similarly the style of same teacher may be different from subject to subject. The suggested automata model helps the children with learning disability to overcome their deficiency in particular domain. This model can be application at any level of education.

But it is recommended to apply from the early education to avoid maximum learning disability.